\documentclass[12pt]{elsarticle}
\usepackage{epsfig}
\usepackage{amssymb}

\usepackage{graphicx}
\usepackage{psfig}

\usepackage{ifthen}
\newboolean{SimonsSetup}
\setboolean{SimonsSetup}{true}
\usepackage{verbatim}
\usepackage{moreverb}

\newcommand{\python}{\mbox{python}}

\newcommand{\half}{\mbox{$\frac{1}{2}$}}

\def\apgt{\,{\raise-.5ex\hbox{$\buildrel>\over\sim$}}\ } 
\def\aplt{\,{\raise-.5ex\hbox{$\buildrel<\over\sim$}}\ } 
\def\lt{\,{\raise-.5ex\hbox{$\buildrel>$}}\ } 
\def\gt{\,{\raise-.5ex\hbox{$\buildrel<$}}\ }

\newcommand{\amusemanager}{manager}
\newcommand{\communitymodule}{community module}
\newcommand{\interface}{communication}
\newcommand{\proxy}{proxy}
\newcommand{\partner}{partner}
\def\aap{\ {A\&A}\ }

\def\aj{\ {AJ}\ }

\def\apj{\ {ApJ}\ }
\def\apjl{\ {ApJL}\ }
\def\apjs{\ {ApJS}\ }

\def\mnras{\ {MNRAS}\ }
\def\nat{\ {Nat}\ }

\def\pasj{\ {Publ. Astr. Soc. Japan}\ }

\begin{document}

\begin{frontmatter}

%BORG: 
\title{Multi-physics simulations using a hierarchical interchangeable
  software interface}

\author{Simon F. Portegies Zwart$^{1}$, 
        Stephen L.W. McMillan$^{2}$,
        Arjen van Elteren$^{1}$,
        F. Inti Pelupessy$^{1}$,
        Nathan de Vries$^{1}$}

\address{
$^{1}$ Sterrewacht Leiden, P.O. Box 9513, 2300 RA Leiden,
                 The Netherlands;\\ 
$^{2}$ Department of Physics, Drexel University, Philadelphia, PA
       19104, USA 
\ead{spz@strw.leidenuniv.nl}
}

\begin{abstract}
  We introduce a general-purpose framework for interconnecting
  scientific simulation programs using a homogeneous, unified
  interface.  Our framework is intrinsically parallel, and
  conveniently separates all component numerical modules in memory.
  This strict separation allows automatic unit conversion, distributed
  execution of modules on different cores within a cluster or grid,
  and orderly recovery from errors.  The framework can be efficiently
  implemented and incurs an acceptable overhead.  In practice, we
  measure the time spent in the framework to be less than 1\% of the
  wall-clock time. Due to the unified structure of the interface,
  incorporating multiple modules addressing the same physics in
  different ways is relatively straightforward.  Different modules may
  be advanced serially or in parallel. Despite initial concerns,
  we have encountered relatively few problems with
  this strict separation between modules, and the results of our
  simulations are consistent with earlier results using more
  traditional monolithic approaches.  This framework provides a
  platform to combine existing simulation codes or develop new
  physical solver codes within a rich ``ecosystem'' of interchangeable
  modules.
\end{abstract}


\end{frontmatter}

\section{Introduction}\label{Sect:Introduction}

Large-scale, high-resolution computer simulations dominate many areas
of theoretical and computational science.  The demand for such
simulations has expanded steadily over the past decade, and is likely
to continue to grow in coming years due to the increase in
the volume, precision, and dynamic range of experimental data, as well
as the widening spectral coverage of observations and laboratory
experiments.  Simulations are often used to mine and understand large
observational and experimental datasets, and the quality of these
simulations must keep pace with the increasingly high quality of
experimental data.

In our own specialized discipline of computational astrophysics,
numerical simulations have increased dramatically in both scope and
scale over the past four decades.  In the 1970s and 1980s, large-scale
astrophysical simulations generally incorporated ``mono-physics''
solutions---in our case, the sub-disciplines of stellar evolution
\cite{1973MNRAS.163..279E}, gas dynamics \cite{1977MNRAS.181..375G},
and gravitational dynamics \cite{1985mts..conf..377A}.  A decade
later, it became common to study phenomena combining a few
different physics solvers \cite{2000ApJS..131..273F}.  Today's
simulation environments incorporate multiple physical domains, and
their nominal dynamic range often exceeds the standard numerical
precision of available compilers and hardware
\cite{1985mts..conf..377A}.

Recent developments in hardware---in particular the rapidly increasing
availability of multi-core architectures---have led to a surge in
computer performance \cite{GPUGems2}.  With
the volume and quality of experimental data continuously improving,
simulations expanding in scope and scale, and raw computational speed
growing more rapidly than ever before, one might expect commensurate
returns in the scientific results returned.  However, a major
bottleneck in modern computer modeling lies in the software, the
growing complexity of which is evident in the increase in the number
of code lines, the lengthening lists of input parameters, the number
of underlying (and often undocumented) assumptions, and the expanding
range of initial and boundary conditions.

Simulation environments have grown substantially in recent years by
incorporating more detailed interactions among constituent systems,
resulting in the need to incorporate very different physical solvers
into the simulations, but the fundamental design of the underlying
codes has remained largely unchanged since the introduction of
object-oriented programming \cite{1962LISP.book..109M} and
patterns \cite{Patterns1987}.  As a result, maintaining and extending
existing large-scale, multi-physics solvers has become a major
undertaking.  The legacy of design choices made long ago can hamper
further development and expansion of a code, prevent scaling on large
parallel computers, and render maintenance almost impossible.  Even
configuring, compiling, and running such a code has become a complex
endeavor, not for the faint of heart.  It has become increasingly
difficult to reproduce simulation results, and independent code
verification and validation are rarely performed, even though all
researchers agree that computer experiments require the same degree of
reproducibility as is customary in laboratory settings.

We suggest that the root cause of much of this code complexity lies in
the traditional approach to incorporating multi-physics components
into a simulation---namely, solving the equations appropriate to all
components in a single monolithic software suite, often written by a
single researcher or research group.  Such a solution may seem
desirable from the standpoint of consistency and performance, but the
resulting software generally suffers from all of the fundamental
problems just described.  In addition, integration of new components
often requires sweeping redesign and redevelopment of the code.
Writing a general multi-physics application from scratch is a major
undertaking, and the stringent requirements of high-quality,
fault-tolerant scientific simulation software render such code
development by a single author almost impossible.

But why reinvent or re-engineer a monolithic suite of coupled
mono-physics solvers when well-tested applications already exist to
perform many or all of the necessary individual tasks?  In many
scientific communities there is a rich tradition of sharing scientific
software. Many of these programs have been written by experts who have
spent careers developing these codes and using them to conduct a wide
range of numerical experiments.  These packages are generally
developed and maintained independently of one another.  We refer to
them collectively as ``community'' software.  The term is intended to
encompass both ``legacy'' codes that are still maintained but are no
longer under active development, and new codes still under development
to address physical problems of current interest.  Together, community
codes represent an invaluable, if incoherent, resource for
computational science.

Coupling community codes raises new issues not found in monolithic
applications.  Aside from the wide range of physical processes,
underlying equations, and numerical solvers they represent, these
independent codes generally also employ a wide variety of units, input
and output methods, file formats, numerical assumptions, and boundary
conditions.  Their originality and independence are strengths, but the
lack of uniformity can also significantly reduce the ``shelf life'' of
the software.  In addition, directly coupling the very dissimilar
algorithms and data representations used in different community codes
can be a difficult task---almost as complex as rewriting the codes
themselves.  But whatever the internal workings of the codes, they are
designed to represent a given domain of physics, and different codes
may (and do in practice) implement alternate descriptions of the same
physical processes.  This suggests that integrating community codes
should be possible using interfaces based on physical principles.

In this paper we present a comprehensive solution to many of the
problems mentioned above, in the form of a software framework that
combines remote function calls with physically based interfaces, and
implements an object oriented data model, automatic conversion of
units, and a state handling model for the component solvers, including
error recovery mechanisms. Communication between the various solvers
is realized via a centralized message passing framework, under the
overall control of a high-level user interface.
In \S\ref{Sect:Framework}, we name our framework MUSE, the
MUlti-physics Software Environment.  An example of the MUSE framework
is presented in \S\ref{sect:example}.  In \S\ref{Sect:AdvancedMUSE} we
describe a production implementation of AMUSE, the astrophysics MUSE
environment, which supports a wide range of programming languages and
physical environments.

\subsection{An historical perspective on MUSE}

The basic concepts of MUSE are rooted in the earliest development of
multi-scale software in computational astrophysics.  The idea of
combining codes within a flexible framework began with the NEMO
project in 1986 at the Institute for Advanced Study (IAS) in Princeton
\citep{1995ASPC...77..398T,2010ascl.soft10051B}.  NEMO was (and still
is) aimed primarily at collisionless galactic dynamics.  It used a
uniform file structure to communicate data among its component
programs.

The Starlab package \cite{2001MNRAS.321..199P,2010ascl.soft10076H},
begun in 1993 (again at IAS), adopted the NEMO toolbox approach, but
used pipes instead of files for communication among modules.  The goal
of the project was to combine dynamics and stellar and binary
evolution for studies of collisional systems, such as star clusters.
The stellar/binary evolution code SeBa \citep{1996A&A...309..179P}
was combined with a high-performance gravitational stellar dynamics
simulator.  Because of the heterogeneous nature of the data, not all
tools were aware of all data types (for example, the stellar evolution
tools generally had no inherent knowledge of large-scale gravitational
dynamics).  As a result, the package used an XML-like tagged data
format to ensure that no information was lost in the production
pipeline---unknown particle data were simply passed unchanged from
input to output.

The intellectual parent of (A)MUSE is the MODEST initiative, begun in
2002 at a workshop at the American Museum of Natural History in New
York.  The goal of that workshop was to formalize some of the ideas of
modular software frameworks then circulating in the community into a
coherent system for simulating dense stellar systems.  Originally,
MODEST stood for MOdeling DEnse STellar systems (star clusters and
galactic nuclei).  The name was later expanded, at the suggestion of
Giampolo Piotto (Padova) to Modeling and Observing DEnse STellar
systems.  The MODEST web page can be found at {\tt
http://www.manybody.org/modest}.  Since then, MODEST has gone on to
provide a lively and long-lived forum for discussion of many topics in
astrophysics.  (A)MUSE is in many ways the software component of the
MODEST community.  An early example of MUSE-like code can be found in
the proceedings of the MODEST-1 meeting \cite{2003NewA....8..337H}.

Subsequent MODEST meetings discussed many new ideas for modular
multiphysics applications \citep[e.g.][]{2003NewA....8..605S}.  The
basic MUSE architecture, as described in this paper, was conceived
during the 2005 MODEST-6a workshop in Lund
\citep[Sweden,][]{2006NewA...12..201D}.  The MUSE name, and the first
lines of MUSE code, were created during MODEST-6e in Amsterdam in
2006, and expanded upon over the next 1--2 years.  The ``Noah's Ark''
milestone (meeting our initial goal of having two independent modules
for solving each particular type of physics) was realized in 2007,
during MODEST-7f in Amsterdam and MODEST-7a in Split
\cite[Croatia][]{2009NewA...14..369P}. 
The AMUSE project, short for
for Astrophysics Multi-Purpose Software Environment, a re-engineered
version of MUSE---``MUSE 2.0,'' building on the lessons learned during
the previous 3 years---began at Leiden Observatory in 2009.

\section{The MUSE framework}\label{Sect:Framework}

Each of the problems discussed in \S\ref{Sect:Introduction} could in
principle be addressed by designing an entirely new suite of programs
from scratch.  However, this idealized approach fails to capitalize on
the existing strengths of the field by ignoring the wealth of highly
capable scientific software that has been developed over the last four
or five decades.  We will argue that it is more practical, and
considerably easier, to introduce a generalized interface that
connects existing solvers to a homogeneous and easily extensible
framework.

At first sight, the approach of assimilating existing software into a
larger framework would appear to be a difficult undertaking,
particularly since community software may be written in a wide variety
of languages, such as FORTRAN, C, and C++, and exhibits an enormous
diversity of internal units and data structures.  On the other hand,
such a framework, if properly designed, could be relatively easy to
use, since learning one simulation package is considerably easier than
mastering the idiosyncrasies of many separate programs.  The use of
standard calling procedures can enable even novice researchers quickly
to become acquainted with the environment, and to perform relatively
complicated simulations using it.

To this end, we propose the Multi-physics Software Environment, or
MUSE.  Within MUSE, a user writes a relatively simple script with a
standardized calling sequence to each of the underlying community
codes.  Instructions are communicated through a message-passing
interface to any of a number of spawned community modules, which
respond by performing operations and transferring data back through
the interface to the user script.

\begin{figure}[htbp] 
\psfig{figure=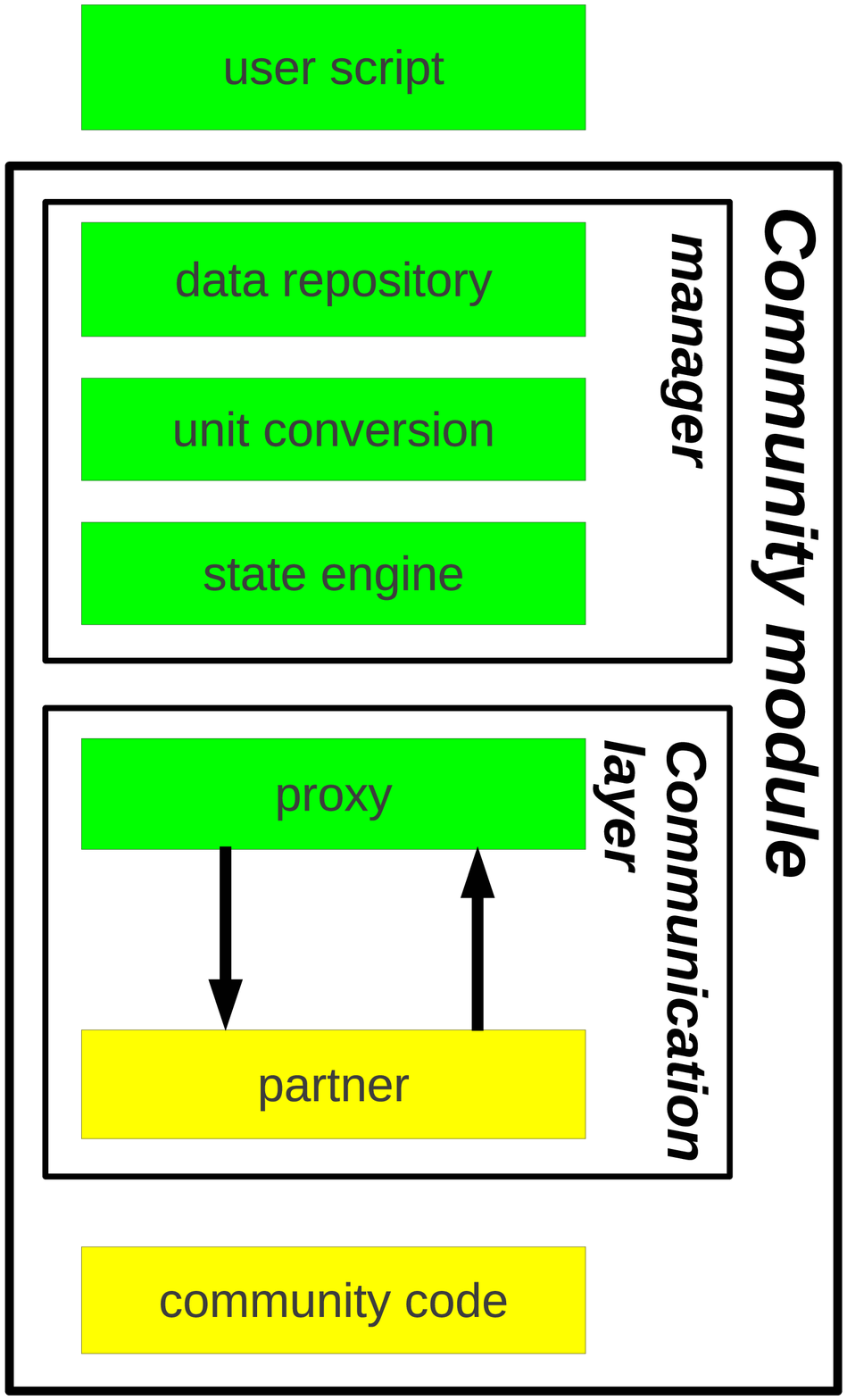,width=\columnwidth}
\caption[]{The MUSE framework architecture is based on a three-layered
  design.  The top layer consists of a user script written by the
  end-user in {\python} (top box).  The {\em{\amusemanager}} is part
  of the {\em{\communitymodule}} and consists of a data repository,
  unit conversion, and a state engine.  The
  {\em{\communitymodule}} also includes the partner-proxy pair in the
  {\interface} layer---a bi-directional message passing
  framework---and a community code (bottom box).
\label{Fig:MUSE_structure}
%http://memory-alpha.org/wiki/Borg_technology
}
\end{figure}

As illustrated in Fig.\,\ref{Fig:MUSE_structure}, the two top-level
components of MUSE:

\begin{itemize}

\item The {\em user script}, which implements a specific physical
  problem or set of problems, in the form of system-provided or
  user-written scripts that serve as the user interface onto the MUSE
  framework.  The coupling between community codes is implemented in
  this layer, with the help of support classes provided by the
  {\em{\amusemanager}}.

\item The {\em{\communitymodule}}, comprising three key elements:

  \begin{enumerate}

  \item The {\em{\amusemanager}}, which provides an object-oriented
    interface onto the {\em{\interface}} layer via a suite of
    system-provided utility functions.  This layer handles unit
    conversion, as discussed below, and also contains the state engine
    and the associated data repository, both of which are required to
    guarantee consistency of data across modules.  The role of this
    layer is generic---it is not specific to any particular problem or
    to any single physical domain.  In \S\ref{Sect:Implementation} we
    discuss an actual implementation of the {\em{\amusemanager}} for
    astrophysical problems.

  \item The {\em{\interface}} layer, which realizes the bi-directional
    communication between the {\em{\amusemanager}} and the community
    code layer.  This is implemented via a {\proxy} and an associated
    {\partner}, which together provide the connection to the community
    code.

  \item The {\em community code} layer, which contains the actual
    community codes and implements control and data management
    operations on them.  Each piece of code in this layer is
    domain-specific, although the code may be designed to be very
    general within its particular physical domain.

  \end{enumerate}

\end{itemize}

\subsection{Choice of programming  languages}
\label{Sect:ComputerLanguage}

We have adopted {\python} as the implementation language for the MUSE
framework and high-level management functions, including the bindings
to the communication interface.  This choice is motivated by
{\python}'s broad acceptance in the scientific community, its object
oriented design, and its ability to allow rapid prototyping, which
shortens the software development cycle and enables easy access to the
community code in the {\communitymodule}, albeit at the cost of
slightly reduced performance.  However, the entire MUSE framework is
organized in such a way that relatively little computer time is
actually spent in the framework itself, as most time is spent in the
community code.  The overhead from {\python} compared to a compiled
high-level language is $\aplt 10$\%, and often much less (see
\S\ref{Sect:Discussion}).

As discussed further in \S\ref{Sect:Communication}, our implementation
of the {\interface} layer in MUSE uses the standard Message Passing
Interface protocol \cite{Lusk96ahigh-performance}, although we also
have an operational implementation using
SmartSockets \cite{2007Maassenetal} via the Ibis
framework \cite{Seinstra_junglecomputing,BalMaassenEtAl10computer}
(see \cite{2012arXiv1203.0321D} for an actual implementation).
Normally MPI is used for communication between compute nodes on
parallel distributed-memory systems.  However, here it is used as the
communication channel between {\em all} processes, whether or not they
reside on the same node as the control script, and whether or not they
are themselves parallel.  The {\proxy} side of the message-passing
framework in the {\interface} layer is implemented in \python, but the
{\partner} side is normally written in the language used in the
community code.  Thus our only real restriction on supported languages
is the requirement that the community code is written in a language
with standard MPI bindings, such as C, C++, or a member of the FORTRAN
family.

\subsection{The {\communitymodule}}

The {\communitymodule} consists of the actual community code and the
bi-directional {\interface} layer enabling low-level MPI communication
between the {\proxy} and the {\partner}.  Each {\communitymodule}
contains a {\python} class of functions dedicated to communication
with the community code.  Direct use of this low-level class from
within a user script is not trivial, because it requires considerable
replication of code dealing with data models, unit conversion, and
exception handling, and is discouraged.  Instead, the MUSE framework
provides the {\amusemanager} layer above the {\communitymodule} layer
to manage the bookkeeping needed to maintain the interface with the
low-level class.

\subsubsection{The community code}

Community applications may be programmed in any computer language that
has bindings to the MPI protocol.  In practice, these codes
exhibit a wide variety of structural properties, model diverse
physical domains, and span broad ranges of temporal and spatial
scales.  In our astrophysical AMUSE implementation (see
\S\ref{Sect:Implementation}), this diversity is exemplified by a suite
of community programs ranging from toy codes to production-quality
high-precision dedicated solvers.  A user script can be developed
quickly using toy codes until the production and data analysis
pipelines are fully developed, then easily switched to production
quality implementations of the physical solvers for actual large-scale
simulations.

In principle, each of the {\communitymodule}s can be used stand alone,
but the main strength of MUSE is in the coupling between them.  Many
of the codes incorporated in our practical implementation are not
written by us, but are publicly available and are amply described in
the literature.

\subsubsection{The {\interface} layer}\label{Sect:Communication}

Bi-directional communication between a running community process and
the {\amusemanager} can in principle be realized using any protocol
for remote interprocess communication.  Our main reasons for choosing
MPI are similar to those for adopting \python---widespread acceptance
in the computational science community and broad support in many
programming languages and on most computing platforms.

The {\interface} layer consists of two main parts, a {\python}
{\proxy} on the {\amusemanager} side of the framework and a {\partner}
on the community code side.  The {\proxy} converts {\python} commands
into MPI messages, transmits them to the {\partner}, then waits for a
response.  The {\partner} waits for and decodes MPI messages from the
{\proxy} into community code commands, executes them, then sends a
reply.  The MPI interface allows the user script and
{\communitymodule}s to operate asynchronously and independently.  As a
practical matter, the detailed coding, decoding, and communications
operations in the {\proxy} and the {\partner} are never hand-coded.
Rather, they are automatically generated as part of the MUSE build
process, from a high-level {\python} description of the
{\communitymodule} interface.

We note that other solutions to linking high-level languages within
{\python} exist---e.g. {\tt swig} and {\tt f2py}, and in fact both
were used in earlier implementations of the MUSE framework
\cite{2009NewA...14..369P}.  However, despite their generality, these
solutions cannot maintain name space independence between
{\communitymodule}s, and in addition are incapable of accommodating
high-performance parallel community codes.  For these reasons, we have
abandoned the standard solutions in favor of our customized,
high-performance MPI alternative.

The use of MPI may seem like overkill in the case of serial operation
on a single computing node (as might well be the case), but it imposes
negligible overhead, and even here it offers significant practical
advantages.  It rigorously separates all community processes in
memory, guaranteeing that multiple instances of a
{\communitymodule} are independent and explicitly avoiding name space
conflicts.  It also ensures that the framework remains thread-safe
when using older community codes that may not have been written with
that concept in mind. 

Despite the initial threshold for its implementation and use, the
relative simplicity of the message-passing framework allows us to
easily couple multiple independent community codes as
{\communitymodule}s, or multiple copies of the same
{\communitymodule}, if desired (see Fig.\,\ref{Fig:MUSE_structure}).
An example of the latter might be to simulate simultaneously two
separate objects of the same type, or the same physics on different
scales, or with different boundary conditions, without having to
modify the data structures in the community code.

As a bonus, the framework naturally accommodates inherently parallel
{\communitymodule}s and allows simultaneous execution of independent
modules from the user script.  Individual {\communitymodule}s can be
offloaded to other processors, in a cluster or on the grid, and can be
run concurrently without requiring any changes in the user script.
This makes the MUSE framework well suited for distributed computing
with a wide diversity of hardware architectures
(see \S\ref{Sect:Discussion}).

\subsubsection{The {\amusemanager}}\label{Sect:Manager}

Between the user script and the communication layer we introduce the
{\amusemanager}, part of the {\communitymodule}.  This part of the
code is visible to the script-writing user and guarantees that all
accessible data are up to date, have the proper format, and are
expressed in the correct units.

The {\amusemanager} layer constructs data models (such as particle
sets or tessellated grids) on top of all functions in the low-level
{\interface} class and checks the error codes they return.  To allow
different modules to work in their preferred systems of units, the
{\amusemanager} incorporates a unit transformation protocol
(see \S\ref{Sect:IntermoduleTransfer}).  It also includes a state
engine to ensure that functions are called in a controlled way, as
many community codes are written with specific assumptions about
calling procedures.  In addition, a data repository is introduced to
guarantee that at least one self-consistent copy of the simulation
data always exists at any given time.  This repository can be
structured in one or more native formats, such as particles, grids,
tessellation, etc, depending on the topology of the data adopted in
the community code.

Each community code has its own internal data, needed for modeling
operations but not routinely exposed to the user script.  However,
some of these data may be needed by other parts of the framework.  To
share data effectively, and to minimize the bookkeeping required to
manage data coherency, the most fundamental parts of the community
data are replicated in the structured repository in the
{\amusemanager}.  The internal data structures differ, but the
repository imposes a standard format.  The repository allows the user
to access {\communitymodule} data from the user script without having
to make individual (and often idiosyncratic) calls to individual
{\communitymodule}s. The repository is updated from the
{\communitymodule} on demand, and is considered to be authoritative
within the script.

The separation of the {\amusemanager} from the {\communitymodule}
realizes a flexibility that allows the user to swap modules and
recompute the same problem using different physics solvers, providing
an independent check of the implementation, validation of the model,
and verification of the results, simply by rerunning the same initial
state using another {\communitymodule}.  Potentially even more
interesting is the possibility of solving some parts of a problem with
one solver and others with a different solver of the same type,
combining the results at a later stage to complete the solution.
While studying the general behavior of a simulation, perhaps to test a
hypothesis or guide our physical intuition, we may elect to use
computationally cheaper physics solvers, then switch to more robust,
but computationally more expensive, modules in a production run.
Alternatively, we might choose adopt less accurate solvers for
physical domains that are not deemed crucial to capturing the correct
overall behavior.

Another advantage of separating of the {\amusemanager} from the
{\communitymodule} is the possibility of combining existing
{\communitymodule}s into new {\communitymodule}s which can themselves
be coupled via the {\amusemanager}, building a hierarchical
environment for simulating complex physical systems with multiple
components.

\subsection{The user script}

The MUSE framework is controlled by the user via {\python} scripts
(see \S\ref{sect:example} and \S\ref{Sect:Implementation}).  The main
tasks of these scripts are to identify and spawn {\communitymodule}s,
control their calling sequences, and manage the data flow among them.
The user script and the spawned community code are connected by a
communication channel embedded in the {\communitymodule} and
maintained by the {\amusemanager}. Since the {\communitymodule}s do
not communicate directly with one another, the only communication
channels are between modules and the user script.  The number of
{\communitymodule}s controlled by a user script is effectively
unlimited, and multiple instances of identical modules can be spawned.

The main objective of the user script is to read in or generate an
input model, process these initial conditions through one or more
simulation modules, and subsequently mine and analyze the resultant
data.  In practice the user script will itself be composed of a series
of scripts, each performing one specific functionality.

\subsection{Inter-module data transfer and unit conversion}
\label{Sect:IntermoduleTransfer}

Computer scientists tend to attribute types to parameters and
variables, and generally impose strict rules for the conversion from
one type to another.  In physics, data types are almost never
important, but units are crucial for validating and checking the
consistency of a calculation or simulation.  The lack of coherent unit
handling in programming environments is notorious, and can cause
significant problems in multi-physics simulations, particularly for
inexperienced researchers.

In many cases, a researcher can define a set of dimensionless
variables applicable to a specific simulation, within the confines of
a set of rather strict assumptions.  For a mono-physics simulation,
the consistent use of such variables is generally clear to most users.
However, as soon as an expert from another field attempts to interpret
the results, or when the output of one simulation is used by another
program, units must be restored and the data converted to another
physical system for further processing and interpretation.

The absence of units in scientific software raises two important
problems.  The first is the loss of an independent consistency check
for theoretical calculations.  The second is the expert knowledge and
intuition required to manage dimensionless variables or otherwise
unfamiliar units.  Few would recognize $4.303 \times 10^{43}$ as the
radius of the Sun in units of the Planck length, even though it is not
completely inconceivable that an astronomer might use such units in a
simulation. To address these issues, MUSE incorporates a
unit-conversion module as part of the {\amusemanager}
(see \S\,\ref{Sect:Manager}) to guarantee that all communications
within the top-level user script are performed in the proper units. In
order to prevent unit checking and conversion from becoming a
performance bottleneck, we adopt lazy evaluation, performing unit
conversion only when explicitly required.

%\subsection{A simple implementation}\label{sect:example}

\section{A simple implementation}\label{sect:example}

The above description of MUSE is somewhat abstract.  Here we present a
simple example---the calculation of the orbital period of a binary
star in a circular orbit.  This is a straightforward astrophysical
calculation that serves to illustrate the basic operation of the
framework.  The community code that calculates the orbital period is
presented in Fig.\,\ref{Source:community.c}.

\begin{figure}[htbp] 
%\ffile{./src/community.c}
\begin{verbatimtab}[8]
#include "community.h"
#include "math.h"

int orbital_period(double orbital_separation, 
		   double total_mass, 
		   double *orbital_period) {
    if (total_mass <= 0.0)
       return -1;
    *orbital_period = sqrt(pow(orbital_separation, 3.)/total_mass) ;
    return 0; 
}
\end{verbatimtab}[8]
\caption{The community code.
         \label{Source:community.c}
}
\end{figure}

The {\partner} requires the definitions of all interface functions in
the community code, which are included from the {\tt community.h}
header file, presented in Fig.\,\ref{Source:community.h}.  (Note that,
as mentioned earlier, in a practical implementation the {\interface}
code is actually machine generated.  The code here is hand-written,
for purposes of exposition.)

\begin{figure}[htbp] 
%\ffile{./src/community.h}
\begin{verbatimtab}[8]
int orbital_period(double orbital_separation, 
		   double total_mass, 
		   double *orbital_period);
\end{verbatimtab}
\caption{The community header file.
         \label{Source:community.h}
}
\end{figure}

\begin{figure}[htbp]  
\psfig{figure=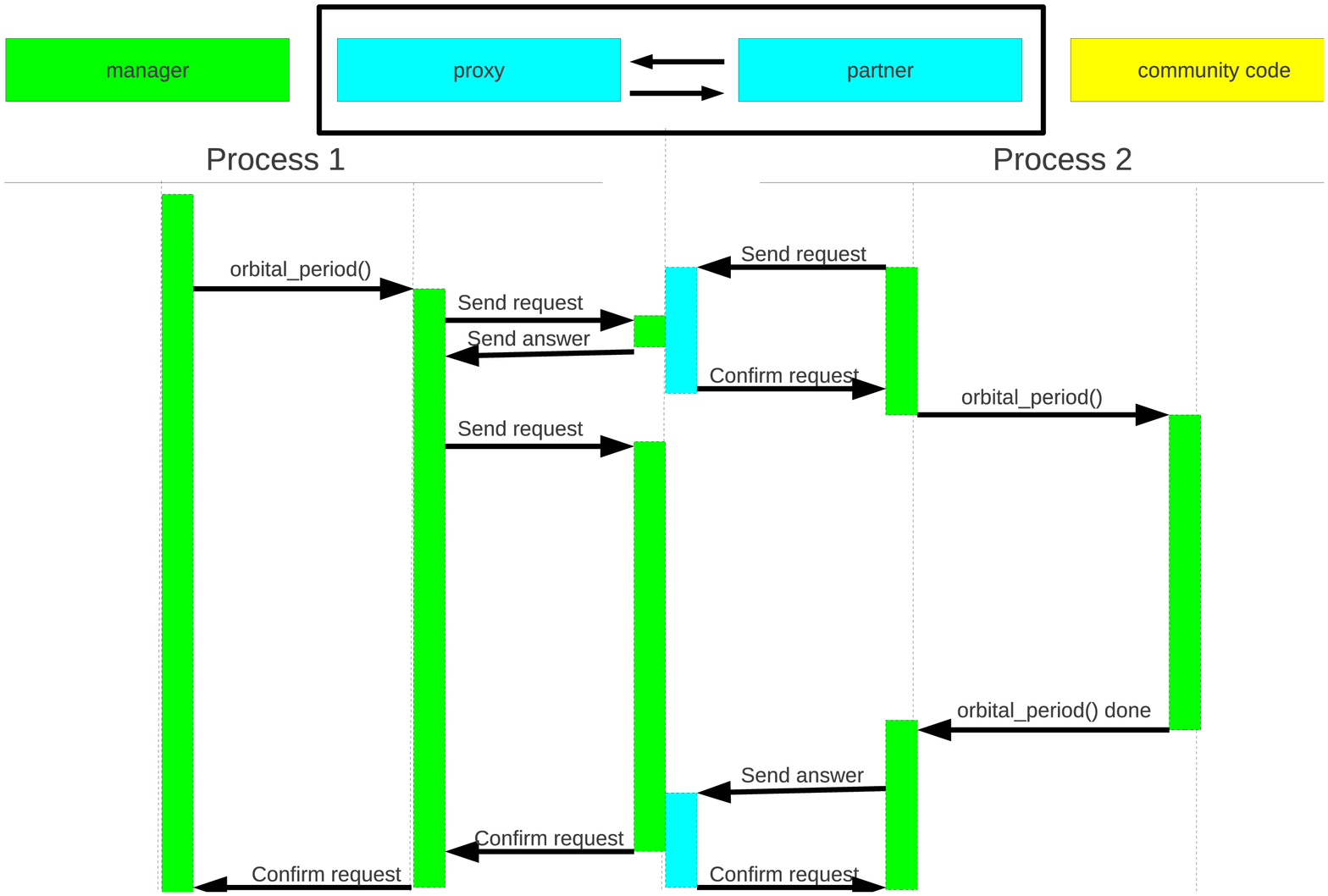,width=\columnwidth}
\caption[]{ The MUSE framework manages the interaction between the
  user script (left) and the community code (right).  When the script
  is started, process 1 spawns the community code and message-passing
  {\partner} as process 2.  Process 2 starts by sending a request for
  instructions, and then waits.  At some later time the user script
  requests the execution of the function {\tt orbital\_period()}.
  This results in a request to the {\interface} layer.  Since process
  2 already has an open request both processes return with the
  confirmation that the request is satisfied.  Process 1 subsequently
  sends a new request to return the resulting data and a confirmation
  that the function {\tt orbital\_period()} has been executed
  correctly.  In the mean time process 2 executes the function,
  returning the requested data and a message to the framework once the
  execution is complete.  \label{Fig:amuse_message_passing} }
\end{figure}

The user script that initializes the required parameters (orbital
separation, total mass of the binary) and computes the orbital period
is presented in Fig.\,\ref{Source:script.py}.  Because of the
simplicity of this example we do not include unit conversion here; we
tacitly assume that the total binary mass is in units of the mass of
the sun, and the orbital separation in astronomical units; the average
orbital separation of Earth around the Sun.  The output orbital period
is expressed in years.  This example illustrates how a set of units
can be convenient within the implicit choices of a community code,
but counterintuitive for researchers from other disciplines. Calls to
the community code are initiated by the user script, sent to the
{\proxy}, received by the {\partner}, and executed by the community
code.  The interaction between these MUSE components is illustrated in
Fig.\,\ref{Fig:amuse_message_passing}.

\begin{figure}[htbp] 
%\ffile{./src/main.py}
\begin{verbatimtab}[8]
import proxy
from optparse import OptionParser

def new_option_parser():
    arguments = OptionParser()
    arguments.add_option("-a", dest="a",type="float",default=1.0)
    arguments.add_option("-M", dest="M",type="float",default=1.0)
    return arguments

def main(a=1.0, M=1.0) :
    community_code = proxy.CodeProxy()
    print "Orbital period: ", community_code.orbital_period(a, M), " years"
    community_code.stop()

if __name__ == '__main__':
    arguments, options  = new_option_parser().parse_args()
    main(**arguments.__dict__)
\end{verbatimtab}
\caption{the main user script.
         \label{Source:script.py}
}
\end{figure}

The {\python} class {\tt{\proxy}} takes care of setting up the community
code, encoding the arguments, sending the message, and decoding the
results.  In an actual MUSE implementation the {\proxy} is split into
two parts, one to translate the arguments into a generic message
object (and translate the returned message into return parameters),
and one to send the message using the communication library (MPI). The
source code listing in Fig.\,\ref{Source:proxy.py} presents a working
example of a proxy.

\begin{figure}[htbp] 
%\ffile{./src/proxy.py}
\begin{verbatimtab}[8]
from mpi4py import MPI
import numpy

class CodeProxy(object):
    def __init__(self):
        self.intercomm = MPI.COMM_SELF.Spawn('./community_code')
    
    def send_function_id(self, function_id):
        self.intercomm.Send([function_id, MPI.CHARACTER], tag=997)
    
    def send_arguments(self, arguments):
        self.intercomm.Send([arguments, MPI.DOUBLE], tag=998)
        
    def recv_error_message(self):
        errorcode_array = numpy.empty(1, dtype='int32')
        self.intercomm.Recv([errorcode_array, MPI.INT], tag = 999)
        errorcode = errorcode_array[0]
        
        if errorcode == -1000:
            raise Exception("Unknown function id received by partner")
        elif errorcode < 0:
            raise Exception("Partner, errorcode is {0}".format(errorcode))
        
    def recv_answer(self, number_of_answers = 1):
        answer_array = numpy.empty(number_of_answers, dtype='float64')
        self.intercomm.Recv([answer_array, MPI.DOUBLE], tag = 996)
        return answer_array
    
    def orbital_period(self, orbital_separation, total_mass):
        arguments = numpy.array([orbital_separation, total_mass])
        self.send_function_id('P')
        self.send_arguments(arguments)
        self.recv_error_message()
        answer = self.recv_answer(1)
        return answer[0]
        
    def stop(self):
        self.send_function_id('q')
        self.recv_error_message()
        self.intercomm.Disconnect()
\end{verbatimtab}[8]
\caption{The {\proxy} part of the {\communitymodule}.
         \label{Source:proxy.py}
}
\end{figure}

Messages sent via MPI are received by the {\partner} code, which
decodes them and executes the actual function calls in the community
code.  Subsequently it encodes and returns the results in a message to
the {\proxy}. The {\partner} code is generally written in the same
language as the community code, in this case C++.  In the source code
listing in Fig.\,\ref{Source:partner.c} we present a working example
of a partner.  Fig.\,\ref{Fig:amuse_message_passing} and the
associated source listing Fig.\,\ref{Source:partner.c} might seem at
first glance to be a rather complicated way to perform some rather simple
message passing. However, this procedure allows us to encapsulate
existing code, rendering it largely independent of the rest of the
framework, allowing the implementation of parallel internal architecture,
and opening the possibility of launching a particular application on a
remote computer. Also, if one of these encapsulated codes stops prematurely,
the framework remains operational, and simply detects the unscheduled
termination of a particular application.  A complete crash of one of
the community modules, however, will likely still cause the framework
to break.

\begin{figure}[htbp] 
%\ffile{./src/partner.c}
\begin{verbatimtab}[8]
#include "community.h"
#include "mpi.h"
#include <mpi.h>

void event_loop() {
    MPI::Intercomm intercomm = MPI::COMM_WORLD.Get_parent();
    int rank = MPI::COMM_WORLD.Get_rank();
    bool continue_run = true;
    
    while (continue_run) {
        char function_id;
        int errorcode = 0;
        intercomm.Recv(&function_id, 1, MPI::CHARACTER, 0, 997);
        
        switch (function_id) {
            case 'P':
                double args[2];
                double answer;
                intercomm.Recv(args, 2, MPI::DOUBLE, 0, 998);
                errorcode = orbital_period(args[0], args[1], &answer);
                if (rank == 0) {
                    intercomm.Send(&errorcode, 1, MPI::INT, 0, 999);
                    if(errorcode == 0) {
                        intercomm.Send(&answer, 1, MPI::DOUBLE, 0, 996);
                    }
                }
                break;
            case 'q':
                intercomm.Send(&errorcode, 1, MPI::INT, 0, 999);
                intercomm.Disconnect();
                continue_run = false;
                break;
            default:
                errorcode = -1000;
                intercomm.Send(&errorcode, 1, MPI::INT, 0, 999);
                break;
        }
    }
}

int main(int argc, char *argv[]) {
    MPI::Init(argc, argv);
    event_loop();
    MPI::Finalize();
}
\end{verbatimtab}
\caption{The {\partner} part of the {\communitymodule}.
         \label{Source:partner.c}
}
\end{figure}

\section{Advanced MUSE}\label{Sect:AdvancedMUSE}

The simple example described in \S\ref{sect:example} does not
demonstrate the full potential of a MUSE framework, and specifically
the wide range of possibilities that arise from coupling community
codes.

Many problems in astrophysics encompass a wide variety of physics.  As
a practical demonstration of MUSE, we present here an open-source
production environment in an astrophysics context, implemented by
coupling community codes designed for gravitational dynamics, stellar
evolution, hydrodynamics, and radiative transfer, the most common
physical domains encountered in astrophysical simulations.  We call
this implementation AMUSE, the Astrophysics MUltiphysics Software
Environment \cite{2011arXiv1110.2785P,2011arXiv1111.3987M}.  

\subsection{AMUSE: a MUSE implementation for astrophysics}
\label{Sect:Implementation}

AMUSE\footnote{see {\tt http://amusecode.org}} is a complete
implementation of MUSE, with a fully functional interface including
automated unit conversion, a structured data repository
(see \S\ref{Sect:Manager}), multi-stepping via operator splitting
(see \S\ref{Sect:Bridge}), and (limited) ability to recover from fatal
errors (see \S\ref{Sect:FailureRecovery}). A wide variety of community
codes is available, including multiple modules for each of the core
domains listed above, and the framework is currently used in a variety
of production simulations.  AMUSE is designed for use by researchers
in many fields of astrophysics, targeted at a wide variety of
problems, but it is currently also used for educational purposes, in
particular for training {MSc.} and {Ph.D.} students.

The development of a MUSE sprang from our desire to simulate a
multi-physics system within a single conceptual framework.  The
fundamental design stems from our earlier endeavour in the development
of the {\tt starlab} software environment \cite{2001MNRAS.321..199P}
in which we combined stellar evolution and gravitational dynamics.
The drawback with starlab was rigid coupling between domains, which
made the framework inflexible and less applicable than desired to a
wider range of practical problems. The AMUSE framework provides us
with a general solution to the problem, allowing us to hierarchically
combine numerical solvers within a single environment to create new
and more capable {\communitymodule}s.  Examples in AMUSE include
coupling a direct N-body algorithm with a hierarchical tree-code, and
combining a smoothed-particle hydrodynamics solver with a larger-scale
grid-based hydrodynamics solver.

Such dynamic coupling has practical applications in resolving short
time scale shocks using a grid-based hydrodynamics solver within a
low-resolution smoothed-particles hydrodynamics environment, or in
changing the evolutionary prescription of a star at run-time---for
example when two stars collide and the standard ``lookup'' description
of their evolution must be replaced by ``live'' integration of the
merger product.  The validation and verification of coupled solvers
remain a delicate issue.  It may well be that the effective strength
of the coupling can only be determined at run-time, and it may require
a thorough study for each application to identify the range of
parameters within which the coupled solver can be applied.

Adding new {\communitymodule}s is straightforward.  The environment
runs efficiently on a single PC, on multi-core architectures, on
supercomputers, and on distributed computers \cite{2012arXiv1203.0321D}.
Given our experience with AMUSE, we are confident that it can be
relatively easily adapted as the basis for a MUSE implementation in
another research field, coupling a different suite of numerical
solvers. The AMUSE source code is freely available from the AMUSE
website.\footnote{{\tt http://www.amusecode.org}}

\begin{figure}[htbp]  
\psfig{figure=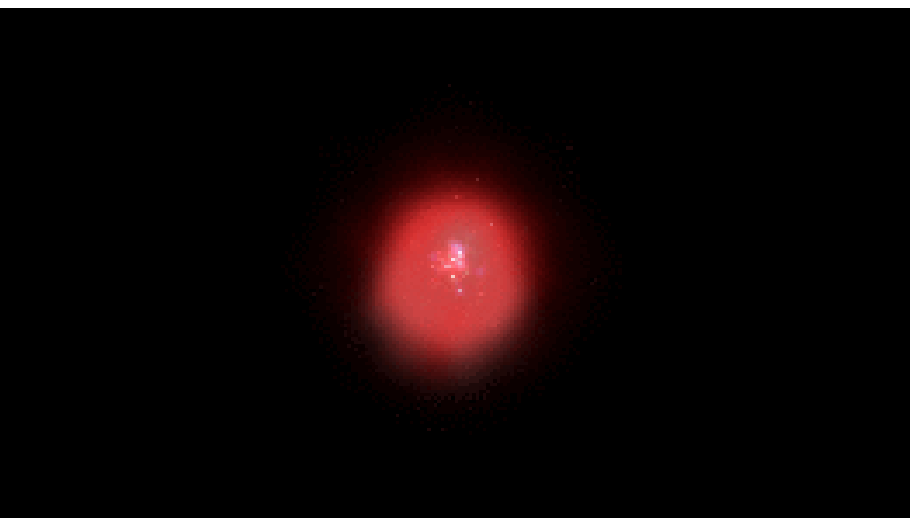,width=0.8\columnwidth}
~\psfig{figure=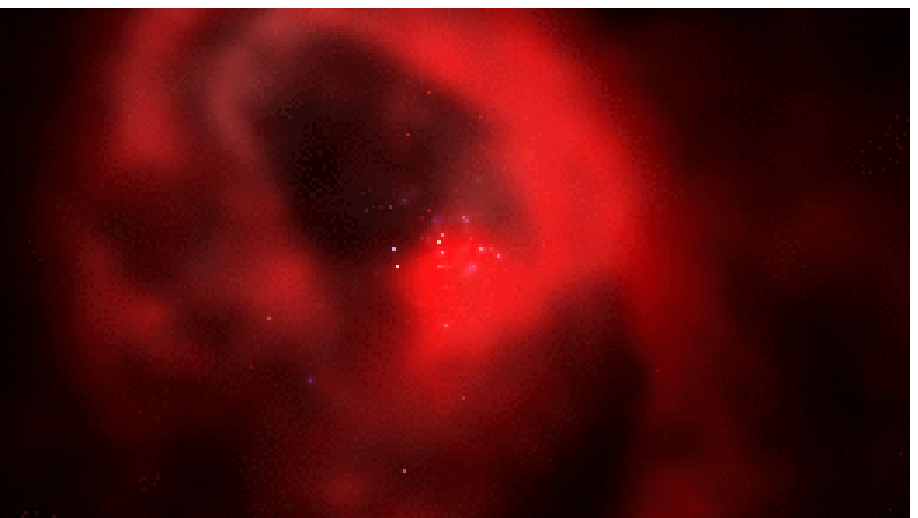,width=0.8\columnwidth}
~\psfig{figure=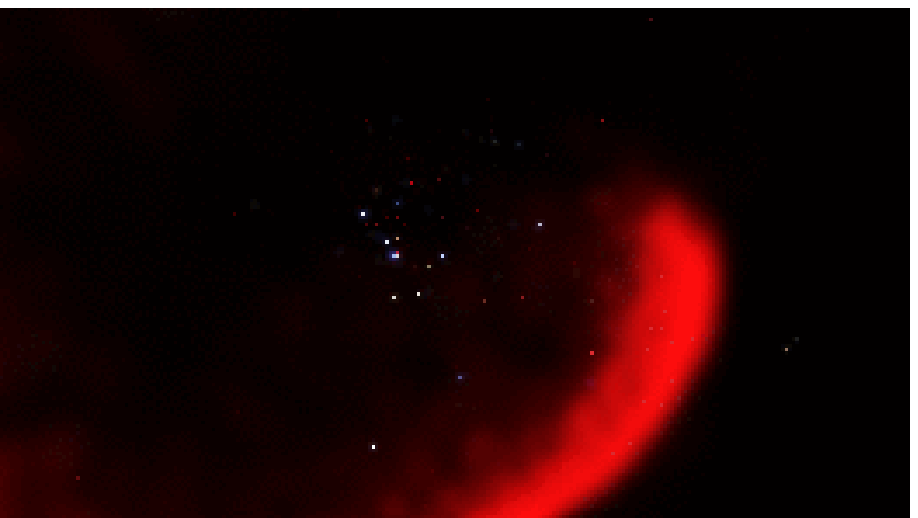,width=0.8\columnwidth}
\caption[]{
           Computer rendering of a simulation of a star cluster (1000
           stars with a Salpeter mass
           function \cite{1955ApJ...121..161S} distributed in a 1\,pc
           Plummer sphere \cite{1911MNRAS..71..460P}) in which the
           mutual gravity and movement of the stars were solved
           self-consistently with the evolution of the stars and the
           hydrodynamics of the protostellar gas, together with the
           stellar outflow in winds and
           supernovae \cite{2012MNRAS.420.1503P}.  The top image
           presents the stars and the gas distribution at the birth of
           the cluster, the second image at an age of about 4\,Myr,
           and the bottom image at about 8\,Myr, just before the last
           residual gas is blown out by the first supernova explosion.
           This 3D visualization is created by Maarten van Meersbergen
           and the animation can be downloaded from {\tt
           http://www.cs.vu.nl/ibis/demos.html} \cite{2012arXiv1203.0321D}.
\label{Fig:PPZ} }
\end{figure}

We have applied AMUSE to a number of interesting problems
incorporating stellar evolution and gravitational dynamics
\cite{2011ApJ...734...55P,2011Sci...334.1380F}, 
and to the combination of stellar evolution and the dynamics of stars
and gas \cite{2012MNRAS.420.1503P}.  The former reference provides an
explanation for the formation of the binary millisecond pulsar
J1903+0327; the latter explores the consequences of early mass loss
from stars in young star clusters by means of stellar winds and
supernovae. In Fig.\,\ref{Fig:PPZ} we present three-dimensional
renderings of several snapshots of the latter calculation.

\subsection{Failure recovery}\label{Sect:FailureRecovery}

Real-world simulation codes crash.  Crashes are inevitable, but
restarting and continuing a simulation afterwards can be a very
delicate procedure, and there is a fine line between science and
nonsense in the resulting data.  In many cases catastrophic failure is
caused by numerical instabilities that are either not understood, in
which case a researcher would like to study them further, or are part
of a natural phenomenon that cannot be modeled in sufficient detail by
the current community code.  Recognizing and understanding code
failure is a time consuming but essential part of the gritty reality
of working with simulation environments.

In many cases, simulation codes are developed with the implicit
assumption that a user has some level of expert knowledge, in lieu of
providing user-friendly error messages, debugging assistance, and
exception handling.  However, for a framework like AMUSE, such an
assumption may pose problems for the run-time behavior and stability
of the system, as well as being confusing and frustrating for the
user.

To some degree, code fatalities can be handled gracefully by the MUSE
{\amusemanager}.  The failure of a particular community module does
not necessarily result in the failure of the entire framework.  As a
response, the framework can return a meaningful error message, or, more
usefully, continuing the simulation using another community code
written for the same problem.  The approach is analogous to
fault-tolerant computing, in which a management process detects node
failure and redirects the processes running on the failed node.

\subsection{Code coupling strategies}

In the AMUSE framework we recognize six distinct strategies for
coupling community modules. Some can be programmed easily by hand,
although they can be labour intensive, while others are enabled by our
implementation of the AMUSE framework.  The examples presented below
are drawn taken from the public version of AMUSE. 

\begin{description}
\item [1 Input/Output coupling:] The loosest type of coupling
  occurs when the result of code {\cal A} generates the initial
  conditions for code {\cal B}.  For example, a Henyey stellar
  evolution code might generate mass, density and temperature profiles
  for a subsequent 3-dimensional hydrodynamical representation of a
  star.

\item [2 One-way coupling:] System {\cal A} interacts with
  (sub)system {\cal B}, but the back-coupling from {\cal B} to {\cal
    A} is negligible.  For example, stellar mass loss due to internal
  nuclear evolution is often important for the global dynamics of a
  star cluster, but the dynamics of the cluster usually does not
  affect the evolution of individual stars (except in the rare case of
  an physical stellar collision).

\item [3 Hierarchical coupling:] Subsystems {\cal A1} and {\cal
    A2} (and possibly more) are embedded in parent system {\cal B},
  which affects their evolution, but the subsystems do not affect the
  parent or each other.  For example, the evolution of cometary Oort
  clouds are strongly influenced by, but are irrelevant to, the larger
  galactic potential.

\item [4 Serial coupling:] A system evolves through distinct
  regimes which need different codes {\cal A}, {\cal B}, {\cal C},
  etc., applied subsequently in an interleaved way, to model them.
  For example, a collision between two stars in a dense star cluster
  may be resolved using one or more specialized hydrodynamics solvers,
  after which the collision product is reinserted into the
  gravitational dynamics code.

\item [5 Interaction coupling:] This type of coupling occurs
  when there is neither a clear temporal or spatial separation between
  systems {\cal A} and {\cal B}.  For example, in the interaction
  between the interstellar medium and the gravitational dynamics of an
  embedded star cluster, both the stellar and the gas dynamics must
  incorporate the combined gravitational potential of the gas and the
  stars.

\item [6 Intrinsic coupling:] This may occur where the physics
  of the problem necessitates a solver that encompasses several types
  of physics simultaneously, and does not allow for temporal or
  spatial separation.  An example is magnetohydrodynamics, where the
  gas dynamics and the electrodynamics are so tightly coupled that
  they cannot be separated.

\end{description}

With the exception of the last, all types of coupling can be efficiently
implemented in AMUSE using single-component solvers.  Many of the
coupling strategies are straightforward in AMUSE, with the exception
of the {interaction} and {intrinsic} coupling types.  For
interaction coupling, the symplectic multi-physics
time-stepping approach originally described in
\cite{2007PASJ...59.1095F} usually works very well.
For intricate intrinsic coupled codes it may still be more efficient
to write a monolithic framework in a single language, rather than
adopt the method proposed here.

\subsubsection{Multi-physics time stepping by operator splitting}
\label{Sect:Bridge}

The relative independence of the various {\communitymodule}s in AMUSE
allows us to combine them in the user script by calling them
consecutively.  This is adequate for many applications, but in other
situations such rigid time stepping is known to have disastrous
consequences for the stability of stiff systems, preventing
convergence of non-linear phenomena \cite{1972SIAM...9..4.603H}.

In some circumstances there is no alternative to simply alternating
between modules in subsequent time steps.  But in others we can write
down the Hamiltonian of the combined solution and integrate this
iteratively, using robust numerical integration schemes.  This
operator splitting approach been demonstrated to work effectively and
efficiently by \cite{2007PASJ...59.1095F}, who adopted the
Verlet-leapfrog algorithm to combine two independent gravitational
N-body solvers.  It has been incorporated into AMUSE for resolving
interactions between gravitational and hydrodynamical
{\communitymodule}s, and is called ``Bridge'' after the introducing
paper \cite{2007PASJ...59.1095F}.

The classical Bridge scheme \cite{2007PASJ...59.1095F} considers a
star cluster orbiting a parent galaxy.  The cluster is integrated
using accurate direct summation of the gravitational forces among all
stars.  Interactions among the stars in the galaxy, and between
galactic and cluster stars, are computed using a hierarchical tree
force evaluation method \cite{1986Natur.324..446B}.

In Bridge, the Hamiltonian of the entire system is divided into two
parts:
\begin{equation}
        H  =  H_A + H_B,
\end{equation}
where $H_A$ is the potential energy of the gravitational interactions
between galaxy particles and the star cluster ($W_{g-c}$):
\begin{equation}
        H_A  =  W_{g-c},
\end{equation}
and $H_B$ is the sum of the total kinetic energy of all particles
($K_g + K_c$) and the potential energy of the star cluster particles
($W_c$) and the galaxy ($W_g$)
\begin{equation}
        H_B  =  K_g + W_g + K_c + W_c \equiv H_g + H_c
\end{equation}

\begin{figure}[htbp]  
\psfig{figure=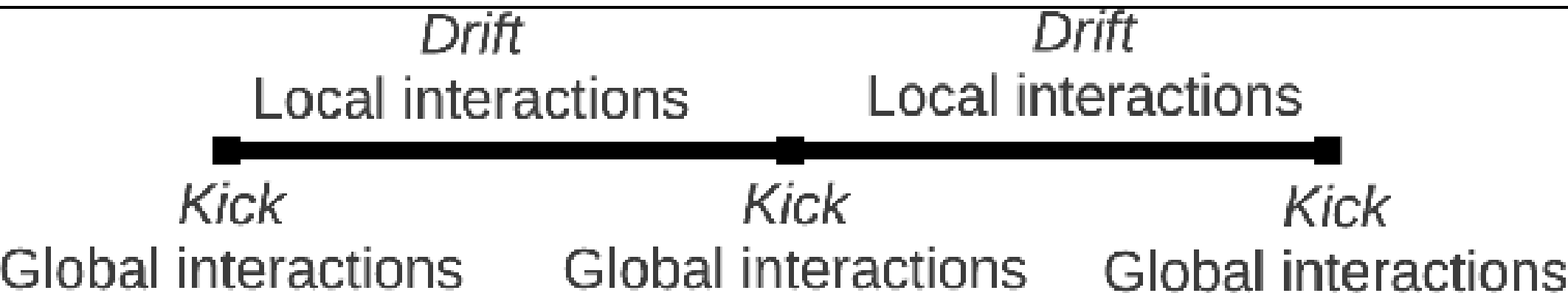,width=\columnwidth}
\caption[]{Schematic kick-drift-kick procedure for the generalized 
           mixed-variable symplectic method \cite{1991AJ....102.1528W}.
           \label{Fig:bridge} }
\end{figure}

The time evolution of any quantity $f$ under this Hamiltonian can then
be written approximately (because we have truncated the formal
solution to just the second-order terms) as:
\begin{equation}
        f^\prime (t+\Delta t) \approx e^{\half \Delta t A} e^{\Delta t B}
	           		     e^{\half \Delta t A} f(t),
\end{equation}
where the operators $A$ and $B$ are defined by $Af = \{f,H_A\}$,
$Bf = \{f,H_B\}$, and $\{.,.\}$ is a Poisson bracket.  The evolution
operator $e^{\Delta t B}$ splits into two independent parts because
$H_B$ consists of two parts without cross terms.  This is the familiar
second-order Leapfrog algorithm.  The evolution can be implemented as
a kick-drift-kick scheme, as illustrated in Fig.\,\ref{Fig:bridge}.

\subsection{Performance}\label{Sect:Performance}

The performance of the AMUSE framework depends on the codes used, the
problem size, the choice of initial conditions and the interactions
among the component parts as defined in the user script. It is
therefore not possible to present a general account of the overhead of
the framework, or the timing of the individual modules used.  However,
in order to provide some understanding of the time spent in the
framework, as opposed to the community modules, we present the results
of two independent performance analysis, one using a suite of 9
gravitational N-body solvers (Figs.\,\ref{Fig:MUSEPerformance}
and\,\ref{Fig:MUSEOverhead}) on three of which we report in more
detail in Fig.\,\ref{Fig:CodeEfficiency}, and analysis of a coupled
hydrodynamics and gravitational dynamics solver in several framework
solutions (Tab.\,\ref{Tab:Timings}).

\subsubsection{Mono-physics solver performance}\label{Sect:MonoPerformance}

\begin{figure}[htbp] 
\psfig{figure=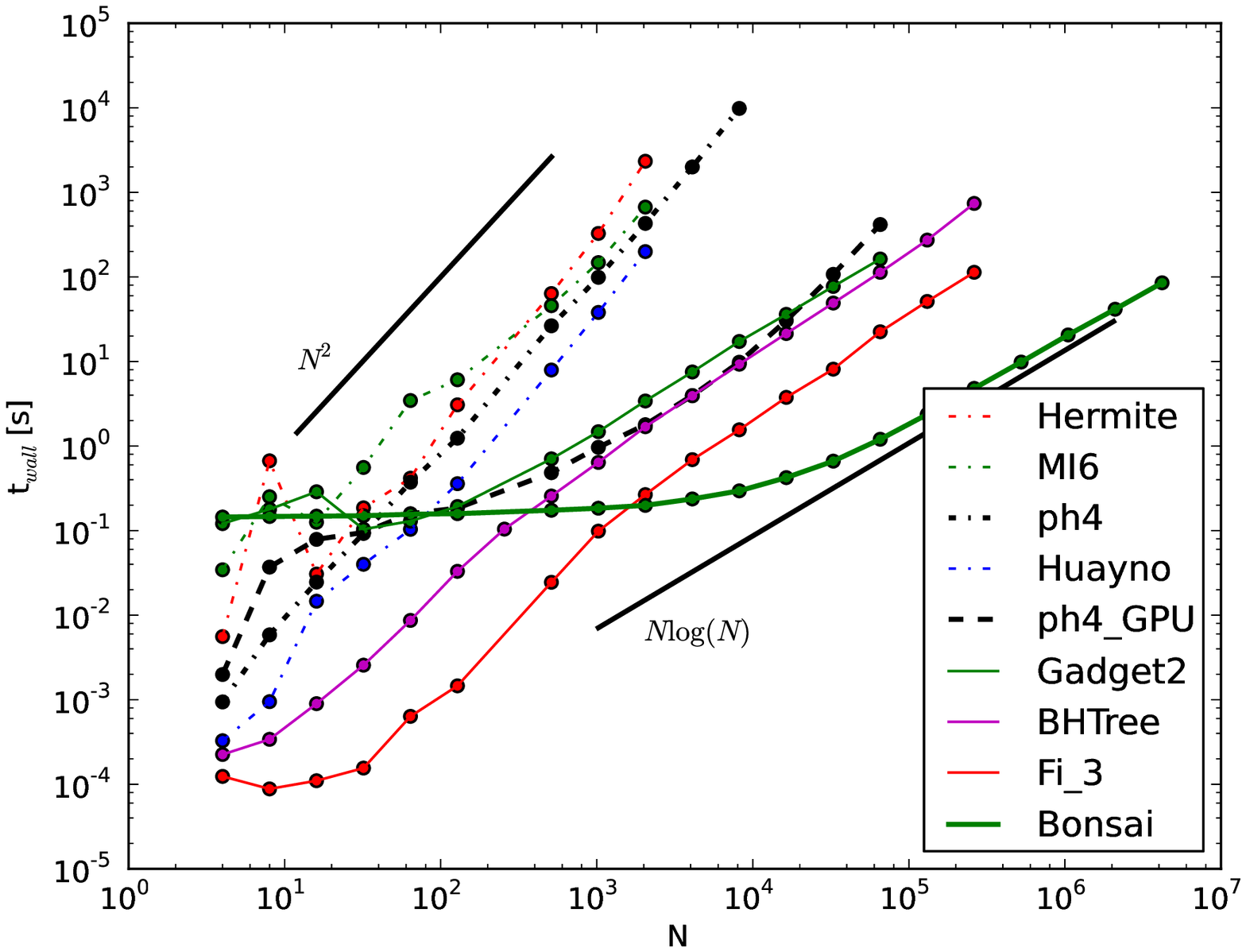,width=\columnwidth}
\caption[]{
Wall clock time for 1 N-body time unit (averaged from running for
10~N-body time units) of a selection of 9 gravitational N-body solvers
in the AMUSE framework.  The direct codes are represented as dashed
curves, the tree codes as solid lines. The scaling for direct N-body
($\propto N^2$) and tree algorithm ($\propto N\log N$) are indicated
with the black solid lines.  The 9 production quality codes include
direct ($N^2$) N-body solvers with a $6^{th}$ order Hermite predictor
corrector integrator
(MI6 \cite{2008NewA...13..498N,2011ApJ...731L...9I}), several $4^{th}$
order Hermite integrators (Hermite \cite{1995ApJ...443L..93H}, Ph4)
semi-symplectic code based on Hamiltonian splitting integrator
(Huayno), and a suite of tree codes
(BHTree \cite{1986Natur.324..446B},
Gadget2 \cite{2005MNRAS.364.1105S},
Bonsai \cite{2012JCoPh.231.2825B}).  For ph4 we also show the GPU
accelerated version, which is realized using the Sapporo library. The
performance of PhiGPU \cite{2007NewA...12..357H} is not shown because
it is almost identical to that of ph4 with GPU support. The reasons
for the similar performance stems from the use of the Sapporo GPU
library for N-body solver, which are used by both codes.  Also
Octgrav \cite{2010ProCS...1.1119G} is omitted, which up to about
$N\sim 10^4$ performs like Bonsai and then follows Fi.  All runs ware
carried out on a 4-core workstation with a Intel Xeon CPU E5620
operating at 2.40GHz and NVIDIA G96 (Quadro FX580) running generic
64-bit Ubuntu Linux kernel 2.6.35-32.
\label{Fig:MUSEPerformance}
}
\end{figure}

In Fig.\,\ref{Fig:MUSEPerformance} we compare the performance of
several N-body solvers in AMUSE. Some of these solvers are GPU
accelerated ({\tt ph4} and {\tt Bonsai} \cite{2012JCoPh.231.2825B}),
others run on multiple cores ({\tt Fi}
\cite{2004A&A...422...55P,2006ApJ...645.1024P}), but most were run on a
single processor even though they could have been run in parallel.
Each calculation was run using standard parameters (double precision,
time step of $dt=2^{-6}$ or an adaptive time step determined by
the community code, an opening angle of $\theta = 75^\circ$ for
tree-codes, and $N^{-1/3}$ softening length).  The initial conditions
were selected from a Plummer \cite{1911MNRAS..71..460P} sphere in
N-body units in which all particles had the same mass. The runs were
performed for 10~N-body time units \cite{1986LNP...267..233H} and
include framework overhead and analysis of the data, but not the
generation of the initial conditions or spawning the community code.

For small $N$ ($\aplt 10$) the performance of all codes saturates,
mainly due to the start-up cost of the AMUSE framework, the
construction of a tree for only a few particles, or communication with
the GPU (which is particularly notable for Bonsai).  For large $N$,
the performance of the direct N-body codes (dashed curves) scales
$\propto N^2$.  The wall-clock time of the tree codes (solid curves)
scale $\propto N\log N$, as expected, but with a rather wide range in
start-up time for small N depending on the particulars of the
implementation. The largest offset is measured for Bonsai; because of
its relatively large start-up time of 0.2\,s, its timing remains
roughly constant until $N\simeq 10^3$, after which it performs
considerably better than the other (tree)codes, which have a smaller
offset but reach their terminal speed at relatively small $N$. It is
in particular due to the use of the GPU that the start-up times for
Bonsai (and also Octgrav, not shown) dominate until quite large $N$,
and terminal speed is not reached until $N\apgt 10^4$.  These
codes require large $N$ to fully benefit from the massive parallelism
of the GPU.

\begin{figure}[htbp] 
\psfig{figure=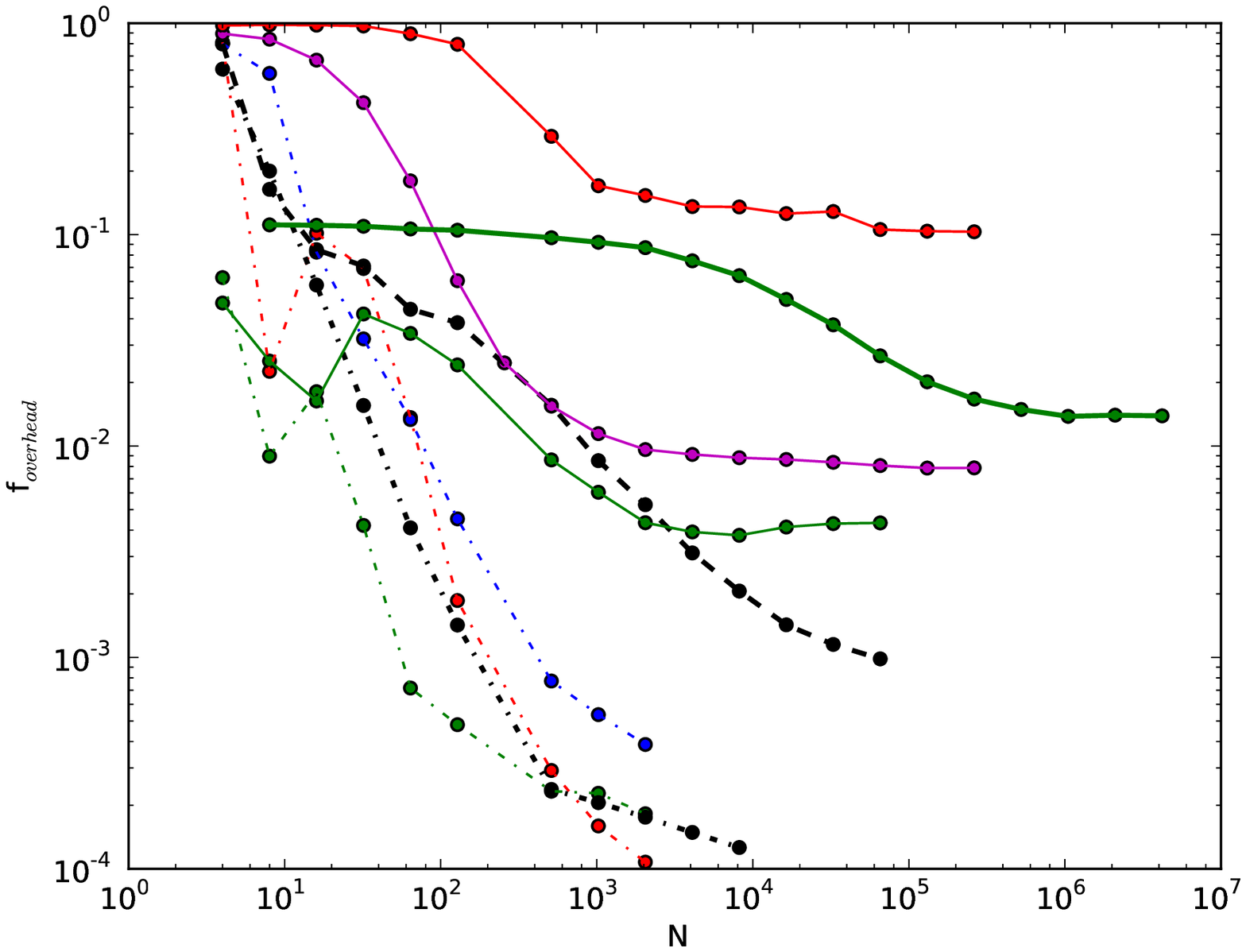,width=\columnwidth}
\caption[]{Overhead (fraction of time spent in the AMUSE framework) 
in a number of N-body simulation codes embedded in the AMUSE
framework.  The line styles are as in Fig.\,\ref{Fig:MUSEPerformance}.
\label{Fig:MUSEOverhead}
}
\end{figure}

In Fig.\,\ref{Fig:MUSEOverhead} we present the fraction of the wall
clock time (in Fig.\,\ref{Fig:MUSEPerformance}) spent in the AMUSE
framework. This figure gives a different perspective from
Fig.\,\ref{Fig:MUSEPerformance}.  The tree codes perform generally
worse in terms of efficiency, mainly because of the relatively small
amount of calculation time spent in the N-body engine compared to the
much more expensive direct N-body codes; this is particularly
notable in Fi and Bonsai. Eventually, each code reaches a terminal
efficiency, the value of which depends on the details of the
implementation. Direct N-body codes tend to converge to an average
overhead of $\aplt 0.1$\% of the runtime, whereas tree codes
reach something like $\sim 1$\% for the fastest code ({\tt Bonsai}).
The particular shape of the various curves depends on the scaling with
$N$ of the work done by the community code and the framework.  We can
draw a distinction between generating the initial conditions ($\mathcal{O}(N)$), starting up the community module ($<\mathcal{O}(N)$),
committing the particle data to the code ($\mathcal{O}(N)$), 
analyzing the data, and the
actual work done in the community code.  The scaling with $N$ of these
last two tasks depends on the implementation --- $\mathcal{O}(N^2)$
for the direct codes and $\mathcal{O}(N\log N)$ for the tree codes.

\begin{figure}[htbp] 
\psfig{figure=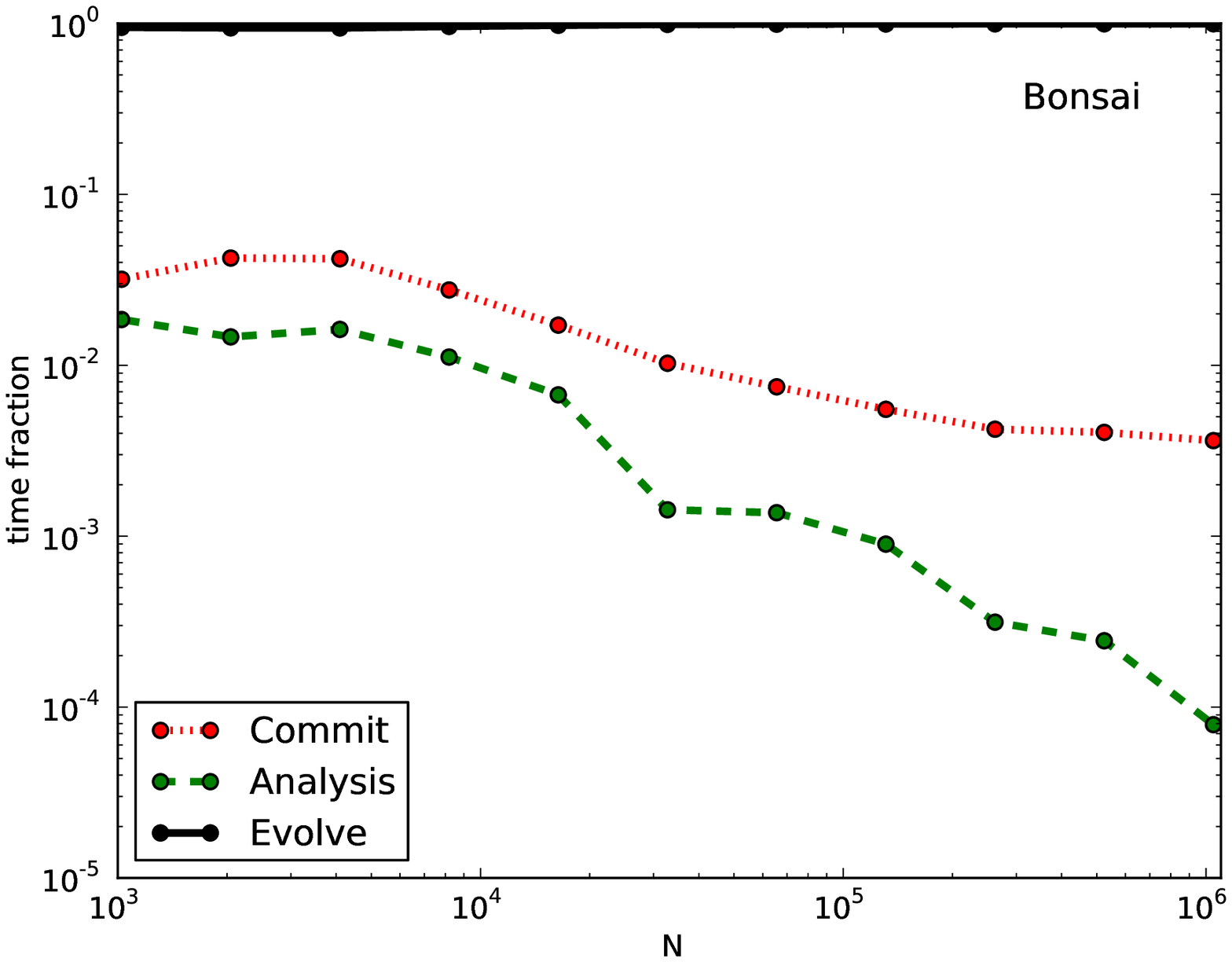,width=0.7\columnwidth}
\psfig{figure=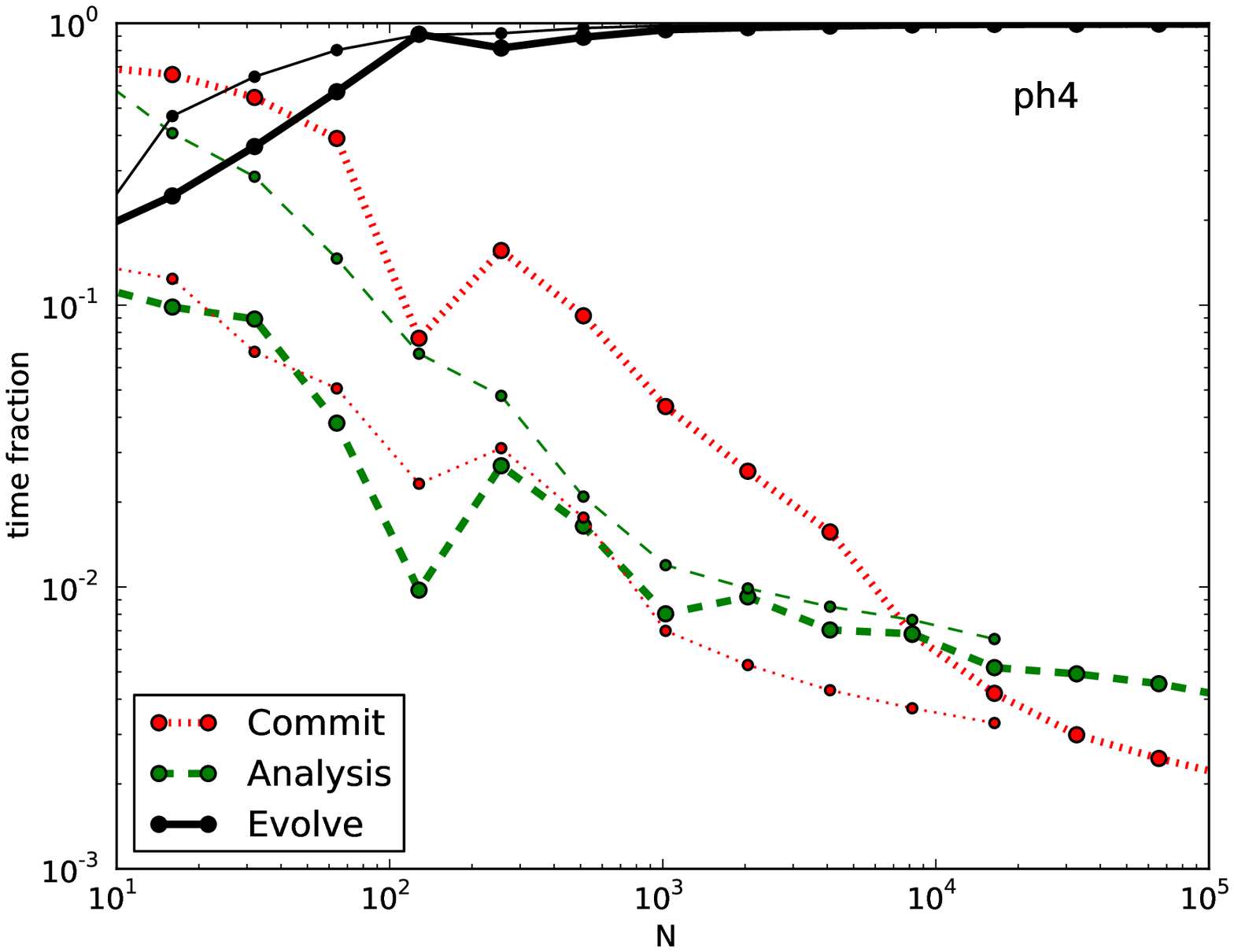,width=0.7\columnwidth}
\caption[]{
Fractional time spent in various parts of the framework as a
function of the number of particles, $N$ for the {\tt Bonsai} tree
code (top panel) and for ph4 with and without GPU support using the
Sapporo library (bottom panel, thick line with GPU, thin line without
GPU).  The top black curve gives time spent to integrate the equations
of motion for a Plummer sphere with equal mass particles without
softening for 1 N-body time unit. The lower curves, indicated in the
legend detail the time for AMUSE to convert units and send the data to
the spawned community code (commit) and the time spent in analyzing
the data (analysis, which in this case also includes calculating the total
kinetic and potential energy of the system).
\label{Fig:CodeEfficiency}
}
\end{figure}

In Fig.\,\ref{Fig:CodeEfficiency} we present a break down of the
wall-clock times for three quite different community codes, but
otherwise identical AMUSE scripts: the Barnes-Hut tree code {\tt
Bonsai}, and the direct code {\tt ph4} with and without GPU support.
We separated the costs for running the community code, committing data
to the code, and data analysis; the last two we call the framework
overhead.  Note that, in practice, the computationally intensive parts
of the analysis calculations are performed by the modules, and not by
the python framework.  Therefore, the measurements of the framework
overhead presented in Fig.\,\ref{Fig:CodeEfficiency} should be
considered as upper limits.  We ignored the cost for start-up
(spawning the community code) and generating the initial conditions,
because these are generally performed only once per run, even though
they may have a time complexity similar to the underlying community
code.  For short runs with few particles these costs may be
substantial, but for production simulations they result in negligible
overhead.

The tree code (top panel in Fig.\,\ref{Fig:CodeEfficiency}) has
$N\log N$ time complexity (see Fig.\,\ref{Fig:MUSEPerformance})
whereas analysis has an $N$ dependency.  For small $N$ the framework
cost may be quite substantial, but for larger simulations these costs
become negligible ($<1$\%).  In the limit of large $N$,
committing the particles has a time complexity similar to a tree
code, but with a much smaller coefficient. For Bonsai, committing
the particles limits the efficiency of the community code to about
99.6\%.  The direct N-body code (bottom panel in
Fig.\,\ref{Fig:CodeEfficiency}) has $N^2$ time complexity (see
Fig.\,\ref{Fig:MUSEPerformance}) and running large $N$ in the
framework is even more favorable in terms of efficiency than
running a tree code. The analysis and commit parts still have the same
scaling with $N$, and for large $N$ less than 1\% of the wall-clock
time is spent in the framework.  The difference between the GPU and
CPU versions of the direct code is most evident in committing the data to
the code, which is relatively slow for the GPU enabled code, and
particularly severe for small $N$.

%Eventually, the performance overhead of the framework will stop
%decreasing, irrespectively of the code used, because some of the other
%tasks of the framework has an $N$-dependence similar to that in the
%community code. For Bonsai that is the commit operation whereas for
%the direct N-body codes it is the data analysis.
% ^ Steve: misleading since analysis generally not part of framework,
% Nathan: even for NlogN codes, commit (N) will always decrease relative to evolve

\subsubsection{Multi-physics performance}\label{Sect:MultiPerformance}

One of the main advantages of AMUSE is the possibility of running
different codes concurrently, with interactions among them, as
discussed in \S\ref{Sect:Bridge}. 
To explore the additional overhead generated in such cases, we
present in Tab.\,\ref{Tab:Timings} timings for a series of
simulations combining gravitational dynamics and hydrodynamics.
The timings listed in the table are determined using three different
codes: the treeSPH code {\tt
Fi} \cite{2004A&A...422...55P,2006ApJ...645.1024P}, the direct
gravitational N-body code {\tt PhiGPU} \cite{2007NewA...12..357H}, and
an analytic static background potential (identified as ``field'' in
the top row of the table).  The treeSPH code is implemented in
FORTRAN90, the direct gravitational $N$-body code is in FORTRAN77, and
the analytic tidal field is implemented in \python.

The simulations were performed for N stars, each of one solar mass, in
a cluster with a Plummer \cite{1911MNRAS..71..460P} initial density
profile and a characteristic radius of 1\,pc.  We simulated the orbits
of cluster stars within a cloud of gas having 9 times as much mass as
in stars, and the same initial density distribution.  The gas content
was simulated using 10 $N$ gas particles.  To treat the direct
gravitational N-body solver {\tt PhiGPU} in the same way as the
tree-code {\tt Fi} for solving the inter-particle gravity, we adopted
a softening length of 0.01\,pc for the stars.  When employing {\tt Fi}
as an SPH code we adopted a smoothing length of 0.05\,pc for the gas
particles.  The N-body time step was 0.01\,N-body time
units \cite{1986LNP...267..233H}, and the interactions between the
stellar- and fluid-dynamical codes were resolved using a Bridge time
step of 0.125\,N-body time units (see
\S\,\ref{Sect:Bridge}, 
but for more details about the simulation
see \cite{2012MNRAS.420.1503P}).

The first column in Tab.\,\ref{Tab:Timings} gives the number of stars
in the simulation.  Subsequent columns give the fraction of the
wall-clock time spent in the AMUSE framework relative to the time
spend in the two production codes.  In the second column, we used {\tt
Fi} for the stars as well as for the gas particles.  No Bridge was
used in this case; rather, the stars were implemented as ``massive
dark particles'' in the SPH code.  For the simulations presented in
the third column we used Bridge to couple the massive dark particles
and the SPH particles, but both the gas dynamics and the gravitational
stellar dynamics were again calculated using Fi as a gravitational
tree-code.  Not surprisingly, the AMUSE overhead using Bridge is larger
than when all calculations are performed entirely within {\tt Fi}
(compare the second and third columns in Tab.\,\ref{Tab:Timings}), but
for $N\apgt 500$, the overhead in AMUSE becomes negligible.  For
$N=1000$ the simulation lasted 769.1\,s\, when using only {\tt Fi},
whereas when we adopted Bridge the wall-clock time was 897.7\,s.  In
production simulations the number of particles will generally greatly
exceed 1000, and we consider the additional cost associated with the
flexibility of using Bridge time well spent.

\begin{table}[htbp]
\begin{center}
\begin{tabular}{crcccccccccc}\hline
N   & Fi    & Fi+Fi& {\tt PhiGPU}+Fi & Fi+field & {\tt PhiGPU}+field \\
10  & 0.1980& 0.563& 0.243          & 0.943    & 0.170 \\
50  &       & 0.129& 0.109          & 0.885    & 0.308 \\
100 & 0.0010& 0.061& 0.057          & 0.701    & 0.322 \\
500 &       & 0.021& 0.021          & 0.200    & 0.377 \\
1000& 0.0003& 0.017& 0.017          & 0.113    & 0.415 \\
5000& 0.0003& 0.016& 0.017          & 0.053    & 0.446 \\
\hline
\hline
\end{tabular}
\caption{\label{Tab:Timings} Relative timing measurements of AMUSE in
  various configurations.  The first column gives the number of star
  particles used in the simulations.  Subsequent columns give the
  fraction of time spent in the AMUSE interface, rather than in the
  module(s).  The relative timings in the second columns are
  performed using Fi as a stand alone module for both the gravity and the
  SPH inside AMUSE.  For the other simulations we adopted a Bridge
  (see \S\,\ref{Sect:Bridge}) interface using two codes, for which we
  adopted {\tt Fi}, {\tt PhiGPU}, and a simple analytic tidal potential
  (field). }
\end{center}
\end{table}

For small $N$ ($\aplt50$), a considerable fraction (up to 83\%) of the
wall-clock time is spent in the AMUSE framework. This fraction is
high, particularly for the combination of {\tt PhiGPU} and the
analytic field, because of the small wall-clock time (less than 10
seconds) of these simulations. The complexity of both codes,
$\mathcal{O}(N^2)$ for {\tt PhiGPU} and $\mathcal{O}(N)$ for the
analytic external field introduces a relatively large overhead in the
{\python} layer compared to the gravity solver, in particular since the
gravitational interactions are calculated using a graphical processing
unit, which for a relatively small number of particles introduces a
rather severe communication bottleneck.  This combination of
unfavorable circumstances cause even the $N=5000$ runs with {\tt
PhiGPU} and the tidal field to spend more than 50\% of the total
wall-clock time in the AMUSE framework.  However, we do not consider
this a serious drawback, since such calculations are generally only
performed for test purposes. In this case it demonstrates that AMUSE
has a range of parameters within which it is efficient, but that there
is also a regime for which AMUSE does not provide optimal performance.

\section{Discussion}\label{Sect:Discussion}

We have described MUSE, a general software framework incorporating a
broad palette of domain-specific solvers, within which scientists can
perform detailed simulations combining different component solvers,
within which scientists can perform detailed simulations using
different component solvers, and AMUSE, an astrophysical
implementation of the concept.  The design goal of the environment is
that it be easy to learn and use, and enable combinations and
comparisons of different numerical solvers without the need to make
major changes in the underlying codes.

One of the hardest, and only partly solved, problems in (A)MUSE is a
general way to convert data structures from one type to another.  In
some cases such conversion can be realized by simply changing the
units, or reorganizing a grid, but in more complex cases, for example,
converting particle distributions to a tessellated grid, and vice
versa, the solution is less clear.  Standard procedures exist for some
such conversions, but it is generally not guaranteed that the solution
in unique.

Some drawbacks of realizing the inter-{\communitymodule} communication
using MPI are the relatively slow response of spawning processes, the
limited flexibility of the communication interface, and the
replication of large data sets in multiple realizations of the same
{\communitymodule}.  Although it is straightforward to spawn multiple
instances of the same code, these processes do not (by design) share
data storage, aside from the data repository in the {\amusemanager}.
For codes that require large data storage, for example look-up tables
for opacities or other common physical constants, the data are
reproduced as many times as the process is spawned.  This limitation
may be overcome by the use of shared data structures on multi-core
machines, or by allowing inter-{\communitymodule} communication via
``tunneling,'' where data moves directly between
{\communitymodule}s, rather than being channeled through the
{\amusemanager}.

The {\python} programming environment is not known for speed, although
this is generally not a problem in AMUSE, since little time is spent
in the user script and the underlying codes are highly optimized,
allowing good overall performance.  Typical monolithic software
environments lose performance by unnecessary calls among domains; in
MUSE this is prevented by the loose coupling between
{\communitymodule}s.  We have experimented with a range of possible
ways to couple the various {\communitymodule}s, and can roughly
quantify the degree to which {\communitymodule}s are coupled.  

MUSE, as described here, is best suited for problems in which the
different physical solvers are relatively weakly coupled.  Weak and
strong coupling may be distinguished by the ratio of the time
intervals on which different {\communitymodule}s are called (see also
\S\ref{Sect:AdvancedMUSE}).  For time step ratios $\aplt1/10$, the
AMUSE implementation gives excellent performance, but if the ratio
approaches unity the MUSE approach becomes progressively more
expensive. It is, however, not clear to what extent a monolithic
solver will give better performance in these strongly coupled cases.
The upshot is that there is a range of problems for which our
implementation works well, but there are surely other interesting
multi-physics problems in astrophysics and elsewhere for which this
separation in scales is not optimal.  However, we have experimented
with more strongly coupled {\communitymodule}s, and have not yet found
it to be a limiting factor in the system design.

\section*{Acknowledgments}

We are grateful to Jeroen B\'edorf, Niels Drost, Michiko Fujii,
Evghenii Gaburov, Stefan Harfst, Piet Hut, Masaki Iwasawa, Jun Makino,
Maarten van Meersbergen, and Alf Whitehead for useful discussions and
for contributing their codes.  SPZ and SMcM thank Sverre Aarseth and
Peter Eggleton for their discussion at IAU Symposium 246 on the
possible assimilation of each other's production software, which
started us thinking seriously about (A)MUSE.  This work was supported
by NWO (grants VICI [\#639.073.803], AMUSE [\#614.061.608] and LGM [\#
612.071.503]), NOVA and the LKBF in the Netherlands, and by NSF grant
AST-0708299 and NASA NNX07AG95G in the U.S.

\end{document}